\documentstyle[epsfig]{aipproc}

\begin{document}
\title{Soft Gamma-ray Repeaters in Clusters of Massive Stars}
\author{I. F\'elix Mirabel$^*$$^{\dagger}$, Ya\"el Fuchs$^*$, and Sylvain Chaty$^{\ddag}$}
\address{$^*$Service d'Astrophysique, CEA Saclay,\\ 
Bat. 709, Orme des Merisiers, 91191 Gif sur Yvette cedex, France\\
$^{\dagger}$Instituto de Astronom\'\i a y F\'\i sica del Espacio,
cc67, suc 28. 1428 Buenos Aires, Argentina\\
$^{\ddag}$Department of Physics and Astronomy, The Open University\\
Walton Hall, Milton Keynes, MK7 6AA, United Kingdom
}
\maketitle

\begin{abstract}
Infrared observations of the environment of the two
Soft Gamma-ray 
Repeaters (SGRs) with the best known locations on the sky 
show that they are associated to clusters 
of massive stars. Observations with ISO revealed that 
SGR~1806-20 is in a cluster of giant massive stars, still
enshrouded in a dense 
cloud of gas and dust \cite{fuchs}. SGR~1900+14 is at the edge of a 
similar cluster that was recently found hidden in the glare of a pair 
of M5 supergiant 
stars \cite{vrba00}. Since none of the stars 
of these clusters has shown in the last years significant flux variations 
in the infrared, 
these two SGRs do not form bound binary systems with massive stars.  
SGR~1806-20 is at only $\sim$ 0.4\,pc, and SGR~1900+14 at $\sim$ 0.8\,pc 
from the centers of their parental star clusters. 
If these SGRs were born with typical neutron star runaway 
velocities of $\sim$ 300\,km.s$^{-1}$, they are 
not older than a few 10$^{3}$ years.  
We propose that SGR~1806-20 and SGR~1900+14 are ideal laboratories 
to study the evolution of supernovae 
explosions inside interstellar bubbles produced by the strong winds 
that prevail in clusters of massive stars.
\end{abstract}

\section*{Introduction}
Neutron stars and stellar mass black holes are the last phase of 
the rapid evolution of the most massive stars, which are known to 
be formed in groups. In this context, it is expected that the most 
recently formed collapsed objects should be found near clusters of 
massive stars, still enshrouded in their placental clouds of gas and dust.   
This should be the case for SGRs, if they indeed are very young 
neutron stars \cite{kouveliotou}.
Among the four SGRs that have been identified with certainty, SGR~1806-20 and 
SGR~1900+14 are the two with the best localizations with precisions 
of a few 
arcsec (Hurley et al., 1999 \cite{hurley99a,hurley99b} and references therein). Both are on the galactic plane 
at distances of $\sim$ 14\,kpc, beyond large columns of 
interstellar material; A$_v$ $\sim$ 30\,mag in front of SGR~1806-20 
\cite{corbel} and A$_v$ $\sim$ 19\,mag in front of SGR~1900+14 
\cite{vrba96}. Because of these large optical obscurations along the line of 
sights, and in the immediate environment of the sources, 
infrared observations are needed to understand their origin and nature.

\section*{Infrared observations}
Mid-infrared (5-18 $\mu$m) observations of the environment of 
SGR~1806-20 were carried out 
with the ISOCAM instrument aboard the Infrared Space Observatory (ISO) 
satellite \cite{fuchs}. By chance, the ISO observations were 
made in two epochs, 11 days before, and 1-4 hours after 
a soft gamma-ray burst detected with the Interplanetary Network 
on 1997 April 14 \cite{hurley99a}. 
 
We also observed\footnote{Based on observations collected at the European 
Southern Observatory, La Silla, Chile under proposal numbers 59.D-0719 and
63.H-0511.} 
SGR~1806-20 in the J\,($1.25 \pm 0.30 \,\mu$m), 
H\,($1.65 \pm 0.30 \,\mu$m) and K$'$\,($2.15\pm 0.32 \,\mu$m) bands 
on 1997 July 19,
and SGR~1900+14 in the J\,($1.25 \pm 0.30 \,\mu$m), 
H\,($1.65 \pm 0.30 \,\mu$m) and Ks\,($2.162\pm 0.275 \,\mu$m) bands 
on 1999 July 25, 
at the European Southern Observatory (ESO), using the IRAC2b camera 
on the ESO/MPI 2.2m telescope for SGR~1806-20, and using the NTT/SOFI for 
SGR~1900+14. 
In the near infrared, SGR~1806-20 was 
monitored by us during the last four years, and SGR~1900+14 by 
Vrba et al. (2000) \cite{vrba00}.

\section*{SGRs in  clusters of massive stars}
\begin{figure}[t!] 
\centerline{\epsfig{file=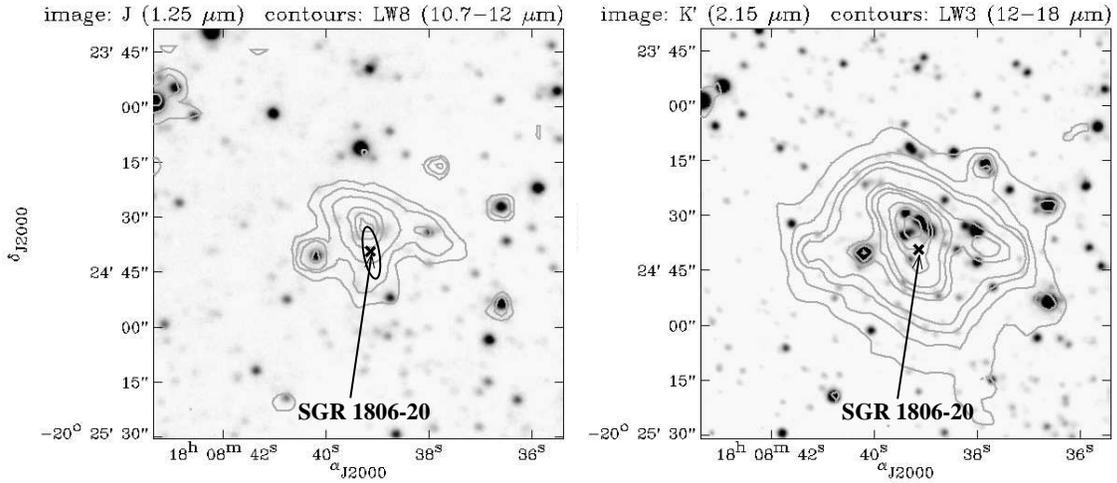,width=15cm}}
\caption{J\,($1.25 \pm 0.32 \,\mu$m) and 
K$'$\,($2.15 \pm 0.32 \,\mu$m) band images of SGR~1806-20 with 0.507''/pixel, 
together with the ISO fluxes ($11.35 \pm 0.65 \,\mu$m) and 
 ($15 \pm 3 \,\mu$m) in contours with 3''/pixel. 
The best fit position of SGR~1806-20 (Hurley et al.\,1999a\,[4])
is marked as a small cross and ellipse. 
These images show that SGR~1806-20   
is $\leq$ 5 arcsec ($\leq$ 0.4 pc at a distance of 14 kpc) 
from the centre of a cluster of hot giant massive 
stars, which are  still partly embedded in their ``placental'' 
cloud of gas and dust.}
\label{fig1}
\end{figure}

\begin{figure}[b!] 
\centerline{\epsfig{file=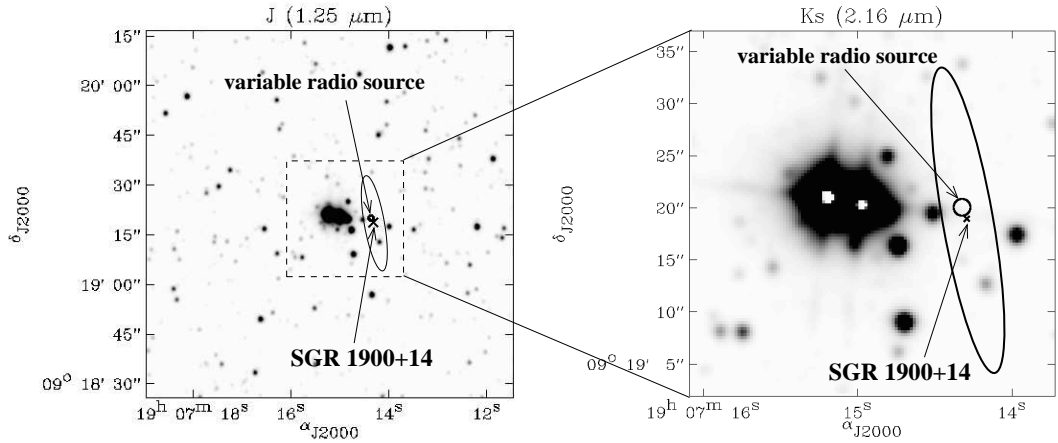,width=15.1cm}}
\caption{ J\,($1.25 \pm 0.32 \,\mu$m) and 
Ks\,($2.162 \pm 0.275 \,\mu$m) band images of SGR~1900+14 with 0.292''/pixel. 
The two white points on the Ks image are due to detector saturation. 
The position of a fading radio counterpart (Frail et al.\,1999\,[10]) 
is indicated by a small circle, and the best fit position of SGR~1900+14 
(Hurley et al.\,1999a\,[4])
is marked as a small cross and ellipse.
SGR~1900+14 is $\sim$\,0.8\,pc from a pair of M5 supergiant stars 
(Vrba et al.\,1996\,[7])
which glare hidde a cluster of stars (Vrba et al.\,2000\,[2]).
SGR~1900+14 is located near the edge of this  
cluster of massive stars.}
\label{fig2}
\end{figure}

The results of the infrared observations of SGR~1806-20 and 
SGR~1900+14 are summarized in Figures \ref{fig1} and \ref{fig2} respectively. 
Figure \ref{fig1} shows a cluster of massive stars deeply embedded in 
a dense cloud of molecular gas and dust. Using the ISO fluxes 
as a calorimeter, Fuchs et al. (1999) \cite{fuchs} show that each of the 
four stars at the centre of the cluster could be equaly, or even more 
luminous than the LBV 
identified in the field by Kulkarni et al. (1995) \cite{kulkarni}.  

van Paradijs et al. (1996) \cite{vanP} reported the possible association 
of SGRs to strong IRAS sources. 
The IRAS fluxes listed in Table \ref{table1} suggest that 
the infrared emission at longer wavelengths detected by IRAS do 
arise in clouds of gas and dust that enshroud these two clusters 
of massive stars.\\

\begin{table}[hb!]
\caption{{\bf Association between SGRs and IRAS sources.} The IRAS
position and 12, 25 and 60 $\mu$m fluxes for 18056-2025 and 19048+0914 
come from van Paradijs et al.\,(1996)\,[9].}
\label{table1}
\begin{tabular}{|c|c|c|c|c|c|c|} 
\tableline \tableline
SGR & SGR Position & IRAS source & IRAS position & \multicolumn{3}{r|}{\, IRAS flux densities (Jy)} \\
    & (J2000) &   & (J2000) & 12 & 25 & 60 $\mu$m \\ 
\tableline \tableline
SGR 1806-20 & $\alpha$ 18 08 39.5 & 18056-2025 & $\alpha$ 18 08 40.4 & 0.98 & 35 & 29 \\
            & $\delta$ -20 24 40   &            & $\delta$ -20 24 41.6  &      &       & \\ \tableline
SGR 1900+14 & $\alpha$ 19 07 14.33 & 19048+0914 & $\alpha$ 19 07 15.3 & 2.5 & 6.3 & 12.3 \\ 
            & $\delta$ +09 19 20.1  &            & $\delta$ +09 19 20.0 &       &       &       \\ \tableline
\end{tabular}
\end{table}

 
\newpage

\section*{CONCLUSIONS}

1) SGR~1806-20 and SGR~1900+14 are associated to clusters of massive stars. 
From ISO observations we find evidence that the cluster associated to 
SGR~1806-20 is enshrouded and heats a dust cloud that appears very bright 
at 12-18\,$\mu$m. Although we did not made ISO observations of SGR~1900+14, 
the latter is as SGR~1806-20, 
a strong IRAS source \cite{vanP}, and very likely 
it is also enshrouded in a dust cloud.\\

2) These SGRs cannot be older than a few 10$^3$ years. At the runaway speeds 
of neutron stars this is the time required to have moved away from the 
centers of their parental clusters of stars.\\

3) J, H and K$'$~bands observations of the massive stars close to the 
SGRs positions  
show no significant flux variations \cite{fuchs,vrba00}.
Therefore, these SGRs do not form bound binary systems 
with any of these massive stars. \\

4) There is strong excess emission 
at 12-18\,$\mu$m associated to SGR~1806-20. However, there is no evidence of 
heating by the high energy SGR activity, although observations were made only 
2 hours after a soft gamma-ray burst reported by Hurley et al. (1999) \cite{hurley99a}.\\

\section*{ACKNOWLEDGMENTS}
The authors are grateful to F.J. Vrba for communicating his results 
on SGR~1900+14 prior to publication.

\end{document}